\documentclass[a4paper]{article}
\usepackage{epsfig}
\begin{document}
\newcommand{\be}{\begin{equation}}
\newcommand{\ee}{\end{equation}}
\newcommand{\bea}{\begin{eqnarray}}
\newcommand{\eea}{\end{eqnarray}}
\newcommand{\nn}{\nonumber \\}
\newcommand{\de}{{\rm d}}
\newcommand{\ie}{{\rm i}}
\newcommand{\cn}{{\rm cn}}
\newcommand{\dn}{{\rm dn}}
\newcommand{\sn}{{\rm sn}}
\newcommand{\frp}[2]{\frac{\partial#1}{\partial#2}}
\newcommand{\ex}[1]{{\rm e}^{#1}}
\newcommand{\xxx}{\rule{2mm}{2mm} }
\newcommand{\du}{\delta u}
\newcommand{\tu}{\tilde u}
\newcommand{\dtu}{\delta\tilde u}
\newcommand{\nicht}[1]{ }
\renewcommand{\Re}{{\rm Re}\,}
\renewcommand{\Im}{{\rm Im}\,}
\newfont{\Kapfont}{cmbx10 scaled 1728}

\vspace*{1cm}
\begin{center}
{\Kapfont Floating Bodies of Equilibrium.\smallskip

Explicit Solution}
\end{center}
\vspace{1cm}

\begin{center}
\bf Franz Wegner, Institut f\"ur Theoretische Physik \\
Ruprecht-Karls-Universit\"at Heidelberg \\
Philosophenweg 19, D-69120 Heidelberg \\
Email: wegner@tphys.uni-heidelberg.de \\
\nicht{\today \ $<$Float/Fl206v2$>$}
\end{center}
\vspace{1cm}

\paragraph*{Abstract} Explicit solutions
of the two-dimensional floating body problem (bodies that can float
in all positions) for relative density $\rho\not=\frac 12$
and of the tire track problem (tire tracks of a bicycle, which do not
allow to determine, which way the bicycle went) are given, which differ
from circles. Starting point is the differential equation given in
\cite{WegnerII,Wegner}.

\section{Introduction}

In this paper explicit solutions are given to the two-dimensional version
of the floating body problem asked by Stanislaw Ulam in the Scottish
Book\cite{Scottish} (problem 19): Is a sphere the only solid of uniform
density which will float in water in any position? A large class of
two-dimensional cross-sections different from the circle were found for
bodies of relative density $\rho=\frac 12$ by Auerbach\cite{Auerbach}.
Here we address to the case $\rho\not=\frac 12$.

It has been shown\cite{Auerbach}, that in two dimensions such a body has the
property, that any chord dissecting the body in two pieces whose areas are
the fractions $\rho$ and $1-\rho$ of the whole area must have the same
constant length $2\ell$. Two equivalent properties are:
(i) Let us consider two close-by water-lines through the cross-section,
$A_1A_2$ and $B_1B_2$. (We keep the body fixed and assumed the direction
of the gravitational force being rotated.) We put the $x$-axis parallel
to $A_1A_2$. Then the vector $A_1B_1$ is given by $(\de x_1,-l_1\de\phi)$
and $A_2B_2$ by $(\de x_2,l_2\de\phi)$. Constant length $\de\ell=0$ implies
$\de x_2=\de x_1$. Constant areas implies $\de f_1=\frac 12 l_1^2\de\phi
=\de f_2=\frac 12 l_2^2\de\phi$. Thus $l_1=l_2=\ell$. This implies, that
the infinitesimal arcs at the perimeter
$\de u_1=\sqrt{(\de x_1)^2+l_1^2(\de\phi)^2}$ and
$\de u_2=\sqrt{(\de x_2)^2+l_2^2(\de\phi)^2}$
are equal. Thus the part of the perimeter below the water-line is constant.
Off course the same is true for the part above the water-line.
One can conclude the other way round: If a curve
has the property, that if
we move from two fixed points $A_1$ and $A_2$ by constant arcs $u$ along the
perimeter and the length of the chord remains fixed, then also the
areas separated by the chord stay constant. We argue, that since $\ell$ stays
constant we have $\de x_1=\de x_2$. Since $\de u_1=\de u_2$ also $l_1=l_2$ and
thus $\de f_1=\de f_2$.
In the main part of the paper we will use this property. By measuring the arc
from a fixed point of the boundary we introduce the arc parameter $u$. We will
show that for certain differences $2\du$ of the arc parameter the length of the
chord is constant, that is the distance between the point at arc
parameter $u-\du$ and at arc parameter $u+\du$ is constant.
\centerline{\epsfig{file=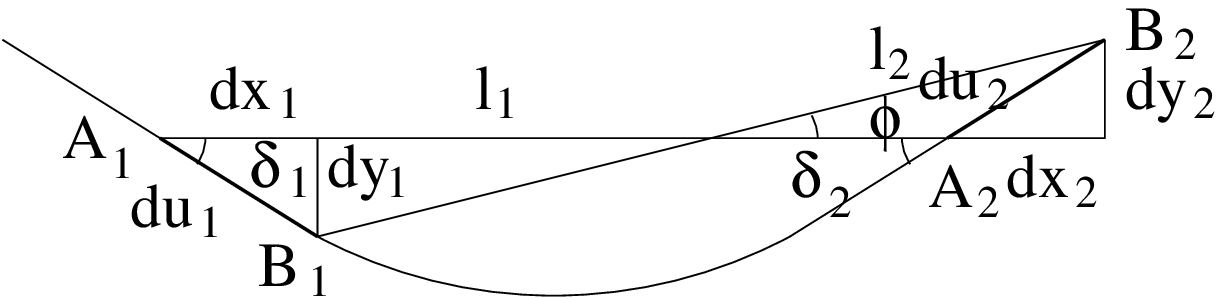,scale=0.5}\hspace{1cm} Fig. 1}

(ii) Another interesting property is, that since $l_1=l_2$, one has
$\de y_1=-\de y_2$,
which implies, that the angles $\delta$ between the tangents and the chord are
equal, $\delta_1=\delta_2$.

The floating body problem is related to the tire track
problem(\cite{Tabachnikov}
and papers cited therein). Let me denote the boundary of the body by $\Gamma$
and the envelope of the waterlines, which consists actually of the
midpoints of the chords, by $\gamma$. Let the distance between the front and
the rear wheel of a bicycle be $\ell$. Then if the front wheel moves along the
curve $\Gamma$ and the rear wheel started on $\gamma$, then it will stay on
$\gamma$. Since however any tangent through a point on $\gamma$ meets $\Gamma$
in the two ends of the chord, the bicycle can move both ways on the same tracks.
The tire track problem consists in finding such curves $\Gamma$, $\gamma$
different from circles, which thus do not allow to determine which way the
bicycle went. It is equivalent to the two-dimensional floating body problem
with the exception that the tire tracks need not close or may wind several
times around some point.

In a recent paper I suggested that one obtains a solution for the boundary of
the floating body from the differential equation
\be
\frac 1{\sqrt{r^2+r^{\prime 2}}} = ar^2+b+cr^{-2}, \label{diffc}
\ee
where $r'=\de r/\de\psi$, with polar coordinates $(r,\psi)$, provided the
resulting curve is sufficiently convex and closed.

By choosing an appropriate condition for the constants $a$, $b$, and $c$
the curves will be closed. They show dihedral $D_p$ symmetry. These curves
have the remarkable property, that they solve the flotation problem for $p-2$
different densities, which implies that there are $[\frac{p-1}2]$ different
chord lenths. (If the boundary is the solution for relative density $\rho$
of the body, then it is also solution for density $1-\rho$ with the same chord
length.) These curves show an even more general property: Let us consider two
copies of these curves (not necessarily closed). Let us choose an arbitrary
point on each curve. Then there exists always an angle by which the two curves
can be rotated against each other, so that the length of the chord between the
points on the two curves stays constant, if we move from the given points by
the same arc. This remarkable property made it possible to obtain the
above-mentioned differential equation by choosing the two points infinitesimally
close \cite{WegnerII,Wegner}.

In \cite{WegnerII,Wegner} I assumed to have proven, that
the solutions have this property, but Serge Tabachnikov kindly
informed me, that this was not correct. What I had shown, was that if for some
chord the angles $\delta$ between the chord and the tangents on the curve obey
$\delta_1=\delta_2$ at $r_1\not=r_2$, that then also the
derivative $\frac{\de(\delta_1-\delta_2)}{\de u}$ vanishes. This is a necessary
condition, but it is not sufficient. That it is not sufficient, can easily be
seen, since from this argument one cannot exclude, that
$\delta_1-\delta_2\propto u^2$. Therefore here the differential
equation will be solved explicitly as a function of the arc parameter $u$, 
and it will be shown, that the curve has the desired property.

First we will introduce an appropriate parametrization. In sect. \ref{radius}
we determine the radius $r$ as a function of the arc $u$, in the next section
the polar angle also as function of $u$. In section \ref{chord} we determine
the length of the chord. Considering the chord between points at arc parameter
$u-\du$ and $u+\du$ for fixed $\du$, we find that the square of
the length of the chord can be expressed as a rational function of the
Weierstrass function $\wp(u/\lambda)$, ($\lambda$ is a constant scale factor).
For special values of $\du$ the residues vanish and thus the chord length
$2\ell$ becomes independent of $u$.
In section \ref{linear} we consider the limit, where the curve oscillates
around a straight line. (They do not constitute solutions of the floating body
problem, but of the tiretrack problem.) There one finds an infinite set of
$\du$s, so that again the
length of the chord between the points of arc parameter $u-\du$ and $u+\du$ is
independent of $u$. In the last section we shortly discuss various shapes as
they appear for example in the papers by Bracho, Montejano and Oliveros
\cite{Oliveros,Bracho} assuming they obey eq. (\ref{diffc}).

\section{Parametrization}

From equation (\ref{diffc}) one obtains
\bea
r^{\prime 2} &=& \frac {r^4}{(ar^4+br^2+c)^2}-r^2, \\
\frac{\de\psi}{\de r} &=& \frac{ar^4+br^2+c}{r\sqrt{r^2-(ar^4+br^2+c)^2}}, \\
\frac{\de\psi}{\de q} &=& \frac{aq^2+bq+c}{2q\sqrt{q-(aq^2+bq+c)^2}} \label{psi}
\eea
with $q=r^2$. If we denote the arc along the curve by $u$ then we obtain
\be
\frac{\de u}{\de\psi} = \sqrt{r^2+r^{\prime 2}}
=\frac 1{ar^2+b+cr^{-2}} = \frac 1{aq+b+cq^{-1}} \label{diffupsi}
\ee
and
\be
\frac{\de u}{\de q} = \frac 1{2\sqrt{q-(aq^2+bq+c)^2}} \label{diffuq}
\ee

With increasing $\psi$ the radius
$r$ oscillates periodically between the largest and the smallest radii $r_>$
and $r_<$, resp. We parametrize these extreme radii by
\be
r_>=r_0(1+\epsilon), \quad r_<=r_0(1-\epsilon).
\ee
We observe, that the polynomial
\be
f(r) := ar^4+br^2+c-r = \frac{r^2}{\sqrt{r^2+r^{\prime2}}} -r
\ee
vanishes at $r=r_>$ and $r=r_<$. Moreover the sum of the zeroes of this
polynomial vanishes. Therefore we introduce a third parameter $\mu$ by writing
\be
f(r) = a(r-r_0(1+\epsilon)) (r-r_0(1-\epsilon)) (r+r_0(1+\ie\mu))
(r+r_0(1-\ie\mu)) \label{fr}
\ee
In order to determine $a$, $b$, and $c$, we first expand the polynomial
\be
f(r) = a(r^4 +r^2r_0^2(-2-\epsilon^2+\mu^2) -2rr_0^3(\epsilon^2+\mu^2)
+r_0^4(1-\epsilon^2)(1+\mu^2)).
\ee
Comparing the coefficient of $r$, $r^2$, and $r^0$ we obtain
\bea
a &=& \frac 1{2r_0^3(\epsilon^2+\mu^2)}, \\
b &=& ar_0^2(-2-\epsilon^2+\mu^2), \\
c &=& ar_0^4(1-\epsilon^2)(1+\mu^2).
\eea
In order to estimate $\mu$ we consider an infinitesimal deformation and
approximate in lowest order in $\epsilon$
\be
r=r_0(1+\epsilon \cos(p\psi))
\ee
Then one has at $r=r_0$
\be
\frac 1{\sqrt{r^2+r^{\prime 2}}}|_{r=r_0}
= \frac 1{r_0}\left(1-\frac{p^2\epsilon^2}2\right)
= \frac 1{r_0^2}f(r_0) +\frac 1{r_0}.
\ee
Substituting $f(r)$ one obtains
\be
\frac 1{r_0^2}f(r_0) = -ar_0^2\epsilon^2(4+\mu^2) = -\frac{r_0p^2\epsilon^2}2
\ee
which yields
\be
\mu^2 = \frac 4{p^2-1}.
\ee
Since one is interested in solutions with $p>1$ one expects real $\mu$.

\section{The radius as function of the arc\label{radius}}

In order to determine the relation between $u$ and $q$ we have to integrate eq.
(\ref{diffuq}). For this purpose we may write
\bea
q-(aq^2+bq+c)^2 = r^2-(ar^4+br^2+c)^2 \nn
=(r-ar^4-br^2-c)(r+ar^4+br^2+c) = -f(r)f(-r) \nn
= -a^2\prod_{i=1}^4 (r_i^2-r^2) = -a^2 \prod_{i=1}^4 (q-r_i^2), \label{pq}
\eea
where we have denoted the four zeroes of $f(r)$ by $r_i$. We introduce the
pa\-ra\-me\-tri\-za\-tion
\be
q=r_0^2\frac{\alpha t+\beta}{t+1}
\ee
with constants $\alpha$ and $\beta$. Then the first two factors are rewritten
\bea
&&(q-r_0^2(1+\epsilon)^2)(q-r_0^2(1-\epsilon)^2) \nn
&=&r_0^4\frac{[(\alpha-(1+\epsilon)^2)t+\beta-(1+\epsilon)^2]
[(\alpha-(1-\epsilon)^2)t+\beta-(1-\epsilon)^2]}{(t+1)^2}.
\eea
We require that the linear term in the numerator vanishes, which yields the
equation
\be
(\alpha-(1+\epsilon^2))(\beta-(1+\epsilon^2))=4\epsilon^2. \label{ab1}
\ee
Secondly we rewrite
\bea
&&(q-r_0^2(1+\ie\mu)^2)(q-r_0^2(1-\ie\mu)^2)
=(q-r_0^2(1-\mu^2))^2+4r_0^4\mu^2 \nn
&=&r_0^4\frac{[(\alpha-1+\mu^2)t+\beta-1+\mu^2]^2+4\mu^2(1+t)^2}{(t+1)^2}.
\eea
The term linear in $t$ in the numerator shall vanish. Thus we require
\be
(\alpha-1+\mu^2)(\beta-1+\mu^2)=-4\mu^2. \label{ab2}
\ee
Equations (\ref{ab1}) and (\ref{ab2}) have the solutions
\bea
\alpha &=& \pm\sqrt{\alpha_+\alpha_-} -1+\frac{\epsilon^2-\mu^2}2 ,\\
\beta &=& \mp\sqrt{\alpha_+\alpha_-} -1+\frac{\epsilon^2-\mu^2}2, \\
\alpha_{\pm} &=& 2+\frac{\epsilon^2+\mu^2}2 \pm 2\epsilon,\nn
\alpha_+\alpha_- &=& \left(2+\frac{\mu^2-\epsilon^2}2\right)^2 +
\epsilon^2\mu^2 \\
&=& \left(2+\frac{\epsilon^2+\mu^2}2\right)^2-4\epsilon^2 \\
&=& \left(2-\frac{\epsilon^2+\mu^2}2\right)^2+4\mu^2.
\eea
Thus we obtain
\be
q-(aq^2+bq+c)^2 = \frac{a^2r_0^8\alpha_+\alpha_-}{(1+t)^4}(A-Bt^2)(C+Dt^2)
\ee
with
\bea
A &=&
-\frac{(\beta-(1+\epsilon)^2)(\beta-(1-\epsilon)^2)}{\sqrt{\alpha_+\alpha_-}}
= \mp(\sqrt{\alpha_+}\pm\sqrt{\alpha_-})^2,\\
B &=&
\frac{(\alpha-(1+\epsilon)^2)(\alpha-(1-\epsilon)^2)}{\sqrt{\alpha_+\alpha_-}}
= \mp(\sqrt{\alpha_+}\mp\sqrt{\alpha_-})^2,\\
C &=& \frac{(\beta-1+\mu^2)^2+4\mu^2}{\sqrt{\alpha_+\alpha_-}}
= 2 (\sqrt{\alpha_+\alpha_-}\pm(2-\frac{\epsilon^2+\mu^2}2)),\\
D &=& \frac{(\alpha-1+\mu^2)^2+4\mu^2}{\sqrt{\alpha_+\alpha_-}}
= 2 (\sqrt{\alpha_+\alpha_-}\mp(2-\frac{\epsilon^2+\mu^2}2)).
\eea
We have incorporated the over-all minus sign in eq. (\ref{pq}) in the
coefficients $A$ and $B$. In the following we choose the lower signs; then all
four coefficients $A$ to $D$ are positive. The differential yields
\be
\de q = r_0^2\frac{\alpha-\beta}{(t+1)^2}\de t
= -2r_0^2\frac{\sqrt{\alpha_+\alpha_-}}{(t+1)^2}\de t.
\ee
Thus we obtain
\be
\de u = \frac{-2r_0(\epsilon^2+\mu^2)\de t}{\sqrt{(A-Bt^2)(C+Dt^2)}}.
\ee
Then $t$ runs between $-\sqrt{A/B}$ and $+\sqrt{A/B}$. Integration yields
\be
-\frac u{\lambda}
= F(\sqrt{\frac{A-Bt^2}A};k), \label{ut}
\ee
with the elliptic integral of the first kind
\be
F(\sin\phi;k) := \int_0^{\sin\phi} \frac{\de t}{\sqrt{(1-t^2)(1-k^2t^2)}}
= F(\phi,k) = \int_0^{\phi} \de\theta \frac 1{\sqrt{1-k^2\sin^2\theta}}.
\ee
(Note the distinction between $F(;)$ and $F(,)$.)
and
\be
k = \sqrt{\frac{AD}{AD+BC}}, \quad
\lambda=\frac{2r_0(\epsilon^2+\mu^2)}{\sqrt{AD+BC}}
\ee
$A-Bt^2$ can be expressed
\be
A-Bt^2 = \frac{\sqrt{4\alpha_+\alpha_-}((1+\epsilon)^2
-\frac q{r_0^2})(\frac q{r_0^2}-(1-\epsilon^2))}
{(\sqrt{\alpha_+\alpha_-}-1+\frac{\epsilon^2-\mu^2}2+\frac q{r_0^2})^2}.
\ee
We obtain
\be
\left.\begin{array}r AD \\ BC \end{array} \right\} =
4(\epsilon^2+\mu^2) \sqrt{\alpha_+\alpha_-}
\pm (8(\epsilon^2-\mu^2)-2(\epsilon^2+\mu^2)^2).
\ee
and thus
\bea
1-2k^2 &=& \frac{BC-AD}{BC+AD}
= \frac 1{\sqrt{\alpha_+\alpha_-}} \left(\frac{2(\mu^2-\epsilon^2)}
{\epsilon^2+\mu^2} +\frac 12 (\epsilon^2+\mu^2) \right), \\
\lambda &=& \frac{r_0\sqrt{\epsilon^2+\mu^2}}
{\sqrt 2 \sqrt[4]{\alpha_+\alpha_-}}.
\eea
From the solution (\ref{ut}) we obtain
\be
t=\sqrt{\frac AB} \cn (\tu,k), \quad
\tu=\frac u{\lambda}
\ee
and
\be
q=r_0^2 \frac{\alpha\sqrt A \cn(\tu)+\beta\sqrt B}{\sqrt A \cn(\tu)+\sqrt B}.
\ee
In the appendix we derive the representation of cn in terms of the Weierstrass
$\wp$-function
\be
\cn(\tu,k) = \frac{\wp(\tu,g_2,g_3)-e_3-\frac 14}
{\wp(\tu,g_2,g_3)-e_3+\frac 14}
\ee
with $g_2$, $g_3$ given in (\ref{g2}, \ref{g3}) and
\be
e_3=\frac{1-2k^2}6.
\ee
Thus we obtain
\be
\frac q{r_0^2} = \frac{2\sqrt{\alpha_+}(1-\epsilon)^2(\wp(\tu)-e_3)
+\frac 12\sqrt{\alpha_-}(1+\epsilon)^2}
{2\sqrt{\alpha_+}(\wp(\tu)-e_3)
+\frac 12\sqrt{\alpha_-}}.
\ee
The arc is measured from $\tu=0$ with $\wp(0)=\infty$, $\cn(0)=1$ at
$q=r^2_0(1-\epsilon)^2$. As $\tu$ increases to
\be
\omega_3 = F(1;k) = F(\frac{\pi}2,k) = 2{\bf K}(k)
\ee
the minimum of $\wp(\tu)$ for real $\tu$ is reached,
$\wp(\omega_3)=e_3$, $\cn(\omega_3)=-1$ and $q=r_0^2(1+\epsilon)^2$.
Then $q$ decreases until $\tu=2\omega_3$, where one period is completed, that is
\be
q(2\omega_3+\tu)=q(\tu).
\ee

Useful references for elliptic integrals and elliptic functions are for example
chapter XIII of \cite{Erdelyi} and chapters 16 to 18 in \cite{Abramowitz}.
The later reference uses $m=k^2$ instead of $k$.
A few relations are listed in the appendix \ref{formulae}.
The elliptic functions are double periodic functions, that is there exist
two periods $2\omega_i$, the ratio of which is not real, so that
\be
\wp(\tu+2\omega_i) = \wp(\tu).
\ee
$2\omega_3$ is a real period of the $\wp$ we use here.

\section{The angle as function of the arc\label{angle}}

We use the expression for $q/r_0^2$ in order to calculate the angle from
(\ref{diffupsi})
\be
\psi(\tu) = \int\de u (aq+b+cq^{-1}).
\ee
Then we have
\bea
\frac q{r_0^2} &=& (1-\epsilon)^2
+\frac{(1-\epsilon)^2(\wp(v_1)-\wp(v_2))}{\wp(\tu)-\wp(v_1)}, \label{q} \\
\frac{r_0^2}q &=& \frac 1{(1-\epsilon)^2}
+\frac{\wp(v_2)-\wp(v_1)}{(1-\epsilon)^2(\wp(\tu)-\wp(v_2))}, \label{invq} \\
\wp(v_1) &=& e_3 -\frac 14\sqrt{\frac{\alpha_-}{\alpha_+}}, \label{wpv1} \\
\wp(v_2) &=& e_3 -\frac{(1+\epsilon)^2}{4(1-\epsilon)^2}
\sqrt{\frac{\alpha_-}{\alpha_+}}. \label{wpv2}
\eea
Since $\wp(v_1)<e_3$ and $\wp(v_2)<e_3$, there are no real solutions $v_1$ and
$v_2$. There are purely imaginary solutions, since $\wp(z)$ is real along the
imaginary axis. $\wp(z)$ starts with $-\infty$ at $z=0$, increases up to $e_3$
at a half-period $\omega'$ and then decreases to $-\infty$ at $z=2\omega'$.
We choose the solution in the interval $0<\Im v_i<\omega'/\ie$. Then
$\wp'(v_i)$ is imaginary and its imaginary part obeys $\Im\wp'(v_i)<0$.
Since the invariants $g_2$ and $g_3$ of $\wp$ are real, one has
$\wp(\tu^*)=\wp^*(\tu)$, where the star indicates the conjugate complex.
In particular for real $\tu$, there is $\wp(\tu-v)=\wp^*(\tu+v)$, since $v$ is
purely imaginary.
\medskip

\epsfig{file=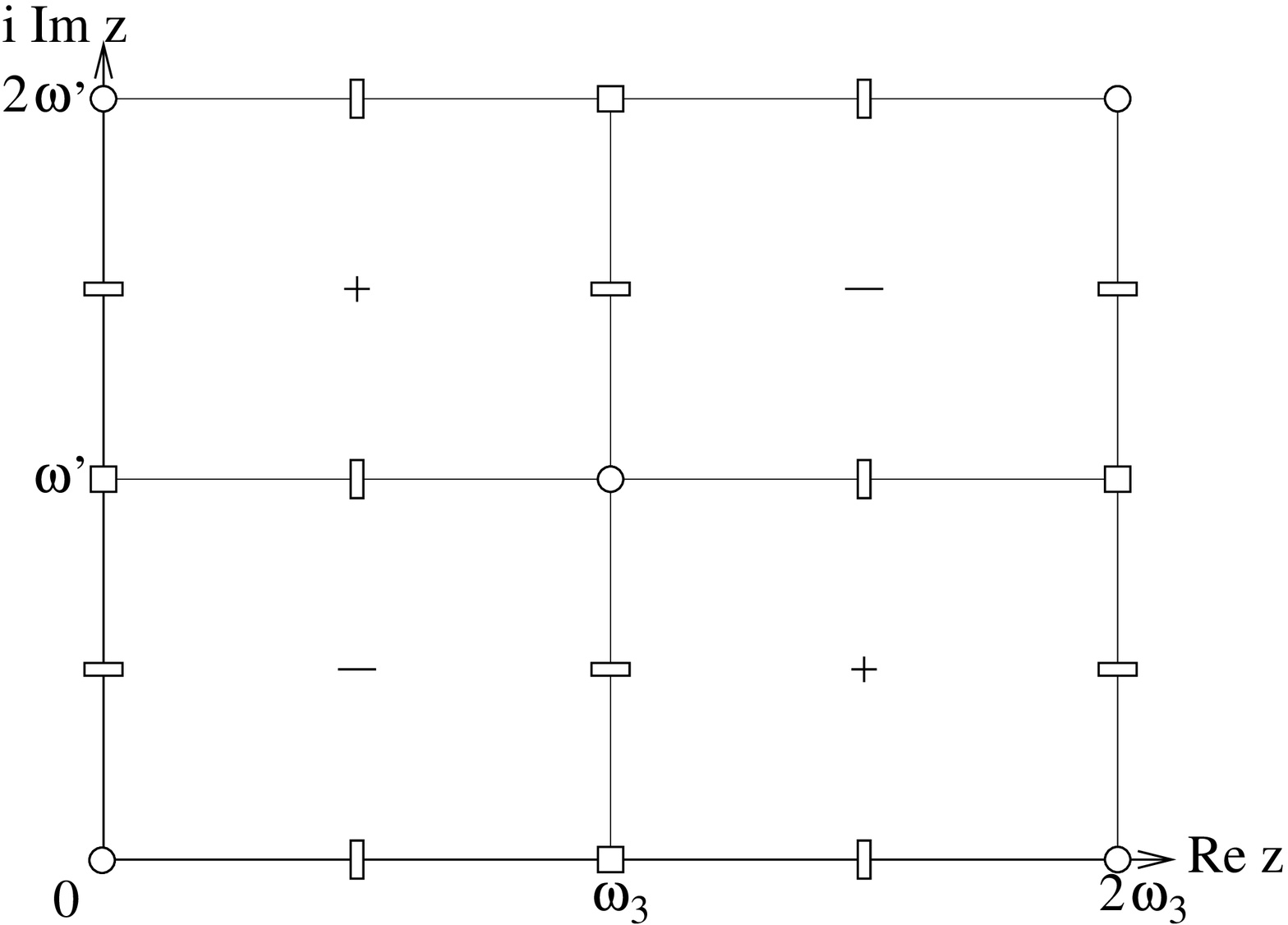,scale=0.4}\hfill
\parbox[b]{3cm}{Fig. 2. Two ele\-men\-ta\-ry cells of pe\-ri\-o\-di\-ci\-ty of
the func\-tion $\wp(z)$.}
\medskip

\newcommand{\rb}[1]{\raisebox{-0.5ex}{\epsfig{file=#1,scale=0.4}}}
The range of the complex plane of $z$ is plotted for $0\le \Re z \le 2\omega_3$,
$0\le \Im z \le 2\omega'/\ie$ in figure 2.
It covers two elementary cells for the double-periodic function $\wp(z)$. The
function is real along the straight lines drawn. For constant $\Im z$
(horizontal lines) it varies between $+\infty$ and $e_3$.
For constant $\Re z$ (vertical lines) it varies between $-\infty$ and $e_3$.
The singularities are at the points indicated by $\rb{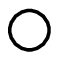}$.
The derivative of the function $\wp(z)$ vanishes at half-periods $\omega$.
$\omega$ is called a half-period, if it is not a period of $\wp$, but $2\omega$
is a period. There are three different half-periods which do
not differ by periods. They are indicated by $\rb{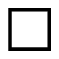}$,
$\rb{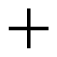}$, and $\rb{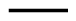}$. Thus one has $\wp'(\rb{weierqu.eps})
= \wp'(\rb{weierpl.eps}) = \wp'(\rb{weiermi.eps})=0$. At these points
the function approaches $\wp(\rb{weierpl.eps})=e_1$,
$\wp(\rb{weiermi.eps})=e_2$, and $\wp(\rb{weierqu.eps})=e_3$. Examples are
$\wp(\omega_1)=\wp(\rb{weierpl.eps})=e_1$ with $2\omega_1=-\omega_3-\omega'$,
$\wp(\omega_2)=\wp(\rb{weiermi.eps})=e_2$ with $2\omega_2=-\omega_3+\omega'$,
$\wp(\omega_3)=\wp(\omega')=\wp(\rb{weierqu.eps})=e_3$. $e_1$ and $e_2$ are
given in the appendix (\ref{e1}, \ref{e2}).
$\omega_1$ and $\omega_2$ lie outside the plotted region.
The locations of the quarter-periods
$\wp(\omega_3/2)=\wp(\rb{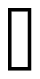})=e_3+1/4$ and
$\wp(\omega'/2)=\wp(\rb{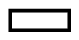})=e_3-1/4$ are also indicated.

The integrals over (\ref{q}, \ref{invq}) are given by
\be
\int\de\tu \frac 1{\wp(\tu)-\wp(v)} = \frac 1{\wp'(v)} \left(2\tu
\zeta(v) + \ln\left(\frac{\sigma(\tu-v)}{\sigma(\tilde u+v)}\right)\right).
\ee
Thus we obtain
\be
\psi(\tu) = \int\de u (aq+b+cq^{-1})
= c_0 u + c_1 \ln\left(\frac{\sigma(\tu-v_1)}{\sigma(\tu+v_1)}\right)
+c_2 \ln\left(\frac{\sigma(\tu-v_2)}{\sigma(\tu+v_2)}\right)
\label{psiu}
\ee
with
\bea
c_0 &=& \frac 1{r_0(1-\epsilon)} +ar_0^2(1-\epsilon)^2
\frac{\wp(v_1)-\wp(v_2)}{\wp'(v_1)} 2 \zeta(v_1) \nn
&+& \frac c{r_0^2(1-\epsilon)^2} \frac{\wp(v_2)-\wp(v_1)}{\wp'(v_2)} 2
\zeta(v_2), \\
c_1 &=& ar_0^2(1-\epsilon)^2 \lambda \frac{\wp(v_1)-\wp(v_2)}{\wp'(v_1)}, \\
c_2 &=& \frac c{r_0^2(1-\epsilon)^2} \lambda
\frac{\wp(v_2)-\wp(v_1)}{\wp'(v_2)}.
\eea
We evaluate $\wp'(v_1)$ and $\wp'(v_2)$ by means of eq. (\ref{wps})
\bea
\wp^{\prime 2}(v_1) &=& -\frac{\sqrt{\alpha_-}\epsilon^2}
{2\alpha^{3/2}_+(\epsilon^2+\mu^2)} \label{wpsv1} \\
\wp'(v_2) &=& \frac{(1+\epsilon)(1+\mu^2)}{(1-\epsilon)^3} \wp'(v_1)
\label{wpsv2}
\eea
The constants $c_i$ evaluate to
\be
c_1 = \frac{\ie}2, \quad c_2 = -\frac{\ie}2 \label{c1c2}
\ee
independent of $\epsilon$ and $\mu$, where we have chosen solutions $v_1$ and $v_2$, for which $\Im\wp'(v_i)<0$. $c_0$ reduces to
\be
c_0 = \frac 1{r_0(1-\epsilon)} - \frac{\zeta(v_1)-\zeta(v_2)}{\ie\lambda}.
\label{c0}
\ee
If we increase $\tu$ by the period $2\omega_3$, then $\psi$ increases by
\bea
\psi_{\rm per} &=& \psi(\tu+2\omega_3) - \psi(\tu)
= 2\omega_3 c_0 \lambda - 2\ie(v_1-v_2)\zeta(\omega_3) \label{psiper} \\
&=& \frac{2\omega_3\lambda}{r_0(1-\epsilon)}
-2\ie \left((v_1-v_2)\zeta(\omega_3)+\omega_3(\zeta(v_2)-\zeta(v_1))\right).
\nonumber
\eea
Thus the curve obeys
\bea
r(\tu+2\omega_3) = r(\tu), && \psi(\tu+2\omega_3)=\psi(\tu)+\psi_{\rm per}, \\
r(-\tu) = r(\tu), && \psi(-\tu) = - \psi(\tu).
\eea

\subsection*{Symmetry}

Some of the equations given above are not explicitly symmetric with respect to
the change of the sign of $\epsilon$. We realize, that $k^2$ and $\lambda$ are
invariant against such a sign change.
However the expression for the angle $\psi_{\rm per}$ of the periodicity of
$r(\psi)$ is not obviously invariant. One can observe however, that the
reversal of the sign of $\epsilon$ corresponds to a change from $v_i$ to
$v'_i=\omega'-v_i$. To see this we write
\be
\wp(v)=e_3-\frac w4.
\ee
and determine
\bea
\wp(\omega'-v) &=& -\wp(\omega')-\wp(v) +\frac{(\wp'(\omega')+\wp'(v))^2}{
(\wp(\omega')-\wp(v))^2} \nn
&=& -2e_3+\frac w4 +\frac{-w(-3e_3w/4+w^2/16+\frac 1{16})}{w^2/4} = e_3-\frac 1{4w},
\eea
where we use eq. (\ref{wps}) and $\wp(\omega')=e_3$, $\wp'(\omega')=0$. Thus under
this transformation $w$ has simply to be replaced by its inverse $1/w$.
Comparing with eqs. (\ref{wpv1}) and (\ref{wpv2}) we see that reversing the sign
of $\epsilon$ corresponds to replacing $v_i$ by $v'_i=\omega'-v_i$. $\Im\wp'(v'_i)$ is negative as $\Im\wp'(v_i)$, but the differences
$\wp(v_1)-\wp(v_2)$ and $\wp(v'_1)-\wp(v'_2)$ have now opposite signs. Therefore
eq. (\ref{psiper}) has to be replaced by
\bea
&& \psi_{\rm per}
= \frac{2\omega_3\lambda}{r_0(1+\epsilon)}
+2\ie \left((v'_1-v'_2)\zeta(\omega_3)+\omega_3(\zeta(v'_2)-\zeta(v'_1))\right)
\nn
&& =  \frac{2\omega_3\lambda}{r_0(1+\epsilon)}
+2\ie \left((v_2-v_1)\zeta(\omega_3)
+\omega_3(\zeta(\omega'-v_2)-\zeta(\omega'-v_1))\right).
\eea
We use now eq. (\ref{zetaadd}) and obtain
\bea
\zeta(\omega'-v_2)-\zeta(\omega'-v_1) = \zeta(v_1)-\zeta(v_2) + \kappa, \\
\kappa = \frac 12 \left(\frac{\wp'(v_2)}{e_3-\wp(v_2)} - \frac{\wp'(v_1)}
{e_3-\wp(v_1)}\right).
\eea
After some algebra one obtains
\be
\kappa = -\ie\frac{\sqrt 2 \epsilon\sqrt{\epsilon^2+\mu^2}}
{\sqrt[4]{\alpha_+\alpha_-}(1-\epsilon^2)}.
\ee
This yields
\be
\psi_{\rm per}
= \frac{2\omega_3\lambda}{r_0(1+\epsilon)}
+ \frac{4\epsilon\omega_3\lambda}{r_0(1-\epsilon^2)}
-2\ie \left((v_1-v_2)\zeta(\omega_3)+\omega_3(\zeta(v_2)-\zeta(v_1))\right),
\ee
which reduces to the expression (\ref{psiper}) for $\psi_{\rm per}$.

Later we will need $\wp(v_1+v_2)$ and $\wp(2v_1)$. By means of eqs. (\ref{wpv1},
\ref{wpv2}) and (\ref{wpsv1}, \ref{wpsv2}) one finds, that
\be
\wp(v_1+v_2) = \wp(2v_1) = e_3
- \frac{\epsilon^2+\mu^2}{8\sqrt{\alpha_+\alpha_-}}. \label{wptwo}
\ee
Since $\wp(v_1+v_2)$ and $\wp(2v_1)$ are equal, we conclude, that
\be
(v_1+v_2) + 2v_1 = 3v_1+v_2 = 2\omega'.
\ee
This implies, that we can express all $v$'s in terms of a variable $\tau$,
\be
v_1=\frac{\omega'}2 + \ie\tau, \quad
v_2=\frac{\omega'}2 -3\ie\tau, \quad
v'_1=\frac{\omega'}2 -\ie\tau, \quad
v'_2=\frac{\omega'}2 +3\ie\tau. \label{tau}
\ee
The sign of $\epsilon$ agrees with the sign of $\tau$.
Thus for positive $\epsilon$ one has
\be
\wp(v_2)<\wp(v'_1)<e_3-1/4<\wp(v_1)<\wp(v'_2)<e_3.
\ee
The derivatives $\wp'(v)$ are purely imaginary.
For positive $\epsilon$ one has
\be
\Im \wp'(v_2) < \Im \wp'(v'_1) < -\frac k2 < \Im \wp'(v_1) < \Im \wp'(v'_2) < 0.
\ee

\section{Constant Chord Length\label{chord}}

Starting out from expression (\ref{psiu}) and using eq. (\ref{sigmaadd})
we obtain for the difference of the angle at $u+\du$ and $u-\du$
\be
\psi(u+\du)-\psi(u-\du) = \chi_0
+ \sum_i c_i
\ln\left(\frac{\wp(\tu)-\wp(\dtu-v_i)}{\wp(\tu)-\wp(\dtu+v_i)}\right),
\ee
where $\chi_0$ is independent of $u$,
\be
\chi_0 = 2c_0\du
+ 2\sum_i c_i\ln\left(\frac{\sigma(\dtu-v_i)}{\sigma(\dtu+v_i)}\right).
\ee
The length $2\ell$ of the chord between these points is given by
\bea
(2\ell)^2 &=& q(u+\du) + q(u-\du) \nn
&&-2\sqrt{q(u+\du) q(u-\du)}
\cos(\psi(u+\du)-\psi(u-\du)) \nn
&=& r_0^2(1-\epsilon)^2 \big(M_1+M_2-\ex{\ie\chi_0}M_3-\ex{-\ie\chi_0}M_4\big)
\eea
with
\bea
M_1 &=& \frac{\Delta(\tu+\dtu,v_2)}{\Delta(\tu+\dtu,v_1)}, \\
M_2 &=& \frac{\Delta(\tu-\dtu,v_2)}{\Delta(\tu-\dtu,v_1)}, \\
M_3 &=& \sqrt{M_1M_2} \ex{\ie(\chi-\chi_0)},\\
M_4 &=& \sqrt{M_1M_2} \ex{-\ie(\chi-\chi_0)}
\eea
and the abbreviation
\be
\Delta(a,b) := \wp(a)-\wp(b).
\ee
By means of the identity (\ref{add2}) derived in the appendix
\be
\Delta(a+b,c)\Delta(a-b,c) =\frac{\Delta^2(a,c)}{\Delta^2(a,b)}
\Delta(a+c,b)\Delta(a-c,b) \label{add3}
\ee
we obtain
\be
\sqrt{M_1M_2}
= \frac{\Delta(\dtu,v_2)}{\Delta(\dtu,v_1)}
\left(\frac{\Delta(\tu,\dtu+v_2)\Delta(\tu,\dtu-v_2)}
{\Delta(\tu,\dtu+v_1)\Delta(\tu,\dtu-v_1)}\right)^{1/2}.
\ee
Together with
\be
\ex{\ie(\chi-\chi_0)} = \left(\frac
{\Delta(\tu,\dtu+v_1)\Delta(\tu,\dtu-v_2)}
{\Delta(\tu,\dtu-v_1)\Delta(\tu,\dtu+v_2)}\right)^{1/2}
\ee
this yields
\bea
M_3 &=& \frac{\Delta(\dtu,v_2)\Delta(\tu,\dtu-v_2)}
{\Delta(\dtu,v_1)\Delta(\tu,\dtu-v_1)}, \\
M_4 &=& \frac{\Delta(\dtu,v_2)\Delta(\tu,\dtu +v_2)}
{\Delta(\dtu,v_1)\Delta(\tu,\dtu+v_1)}.
\eea
On the other hand we may rewrite
\bea
M_1+M_2 &=& \frac{\Delta(\tu+\dtu,v_2)\Delta(\tu-\dtu,v_1)
+\Delta(\tu+\dtu,v_1)\Delta(\tu-\dtu,v_2)}
{\Delta(\tu+\dtu,v_1)\Delta(\tu-\dtu,v_1)} \label{M1+M2} \nn
&=& 1+ \frac{\Delta(\tu+\dtu,v_2)\Delta(\tu-\dtu,v_2)-\Delta^2(v_1,v_2)}
{\Delta(\tu+\dtu,v_1)\Delta(\tu-\dtu,v_1)} \nn
&=& 1+ \frac{\Delta(\tu,\dtu+v_2)\Delta(\tu,\dtu-v_2)\Delta^2(\dtu,v_2)}
{\Delta(\tu,\dtu+v_1)\Delta(\tu,\dtu-v_1)\Delta^2(\dtu,v_1)} \nn
&& -\frac{\Delta^2(\tu,\dtu)\Delta^2(v_1,v_2)}
{\Delta(\tu,\dtu+v_1)\Delta(\tu,\dtu-v_1)\Delta^2(\dtu,v_1)}.
\eea
Thus $(2\ell)^2$ is a rational function of $\wp(\tu)$ with
simple poles at $\wp(\tu)=\wp(\dtu\pm v_1)$,
\bea
\frac{4\ell^2}{r_0^2(1-\epsilon)^2} &=&
1+\frac{\Delta^2(\dtu,v_2)-\Delta^2(v_1,v_2)}{\Delta^2(\dtu,v_1)}
-2\cos(\chi_0)\frac{\Delta(\dtu,v_2)}{\Delta(\dtu,v_1)} \nn
&& +\frac{\nu_1-\nu_4\ex{-\ie\chi_0}}{\Delta(\tu,\dtu+v_1)}
+\frac{\nu_2-\nu_3\ex{\ie\chi_0}}{\Delta(\tu,\dtu-v_1)}
\eea
with
\bea
\nu_1 &=& \frac{z_+}{\Delta(\dtu+v_1,\dtu-v_1)\Delta^2(\dtu,v_1)}, \\
\nu_2 &=& \frac{z_-}{\Delta(\dtu-v_1,\dtu+v_1)\Delta^2(\dtu,v_1)}, \\
\nu_3 &=&
\frac{\Delta(\dtu,v_2)\Delta(\dtu-v_1,\dtu-v_2)}{\Delta(\dtu,v_1)}, \\
\nu_4 &=&
\frac{\Delta(\dtu,v_2)\Delta(\dtu+v_1,\dtu+v_2)}{\Delta(\dtu,v_1)}, \\
z_+ &=& \Delta(\dtu+v_1,\dtu+v_2)\Delta(\dtu+v_1,\dtu-v_2)\Delta^2(\dtu,v_2) \nn
&& -\Delta^2(\dtu+v_1,\dtu)\Delta^2(v_1,v_2), \\
z_- &=& \Delta(\dtu-v_1,\dtu+v_2)\Delta(\dtu-v_1,\dtu-v_2)\Delta^2(\dtu,v_2) \nn
&& -\Delta^2(\dtu-v_1,\dtu)\Delta^2(v_1,v_2).
\eea
If we can choose $\chi_0$ and $\dtu$, so that the two residua vanish, then the
length of the chord does not depend on $u$. It turns out, that the modulus of
all four $\nu$s coincide. First we realize, that
\be
\nu_2=\nu_1^*, \quad \nu_4=\nu_3^*.
\ee
In order to see, that $|\nu_1|=|\nu_4|$ or equivalently $\nu_1\nu_2=\nu_3\nu_4$
we obtain a simplified form for $z_+$ and $z_-$ by applying in the first line
of eq.(\ref{M1+M2}) the identity (\ref{add3}) only to the denominator
\bea
&& M_1+M_2 = \\
&& \frac{\big(\Delta(\tu+\dtu,v_2)\Delta(\tu-\dtu,v_1)
+\Delta(\tu+\dtu,v_1)\Delta(\tu-\dtu,v_2)\big)\Delta^2(\tu,\dtu)}
{\Delta(\tu,\dtu+v_1)\Delta(\tu,\dtu-v_1)\Delta^2(\dtu,v_1)}. \nonumber
\eea
From this expression one reads off the residues $\nu_1$ and $\nu_2$ with
\bea
z_+ &=& \Delta(2\dtu+v_1,v_1)\Delta(v_1,v_2)\Delta^2(\dtu+v_1,\dtu), \\
z_- &=& \Delta(2\dtu-v_1,v_1)\Delta(v_1,v_2)\Delta^2(\dtu-v_1,\dtu).
\eea

To prove $\nu_1\nu_2=\nu_3\nu_4$ we recast the expressions by means of eqs.
(\ref{add5}, \ref{add6})
\bea
\Delta(2\dtu+v_1,v_1)\Delta(2\dtu-v_1,v_1) =
-\frac{\Delta(2\dtu,2v_1)\wp^{\prime 2}(v_1)}{\Delta^2(2\dtu,v_1)}, \\
\Delta(\dtu+v_1,\dtu)\Delta(\dtu-v_1,\dtu)
= \frac{\Delta(2\dtu,v_1)\wp^{\prime 2}(\dtu)}{\Delta^2(\dtu,v_1)}, \\
\Delta(\dtu+v_1,\dtu-v_1) = -\frac{\wp'(\dtu)\wp'(v_1)}{\Delta^2(\dtu,v_1)}
\eea
and obtain
\be
\nu_1\nu_2 = \frac{\Delta(2\dtu,2v_1)\Delta^2(v_1,v_2)\wp^{\prime 2}(\dtu)}
{\Delta^4(\dtu,v_1)}.
\ee
To evaluate $\nu_3\nu_4$ we use eq. (\ref{add4})
\be
\Delta(\dtu+v_1,\dtu+v_2)\Delta(\dtu-v_1,\dtu-v_2)
=\frac{\Delta(2\dtu,v_1+v_2)\Delta^2(v_1,v_2)\wp^{\prime 2}(\dtu)}
{\Delta^2(\dtu,v_1)\Delta^2(\dtu,v_2)}
\ee
and obtain
\be
\nu_3\nu_4 = \frac{\Delta(2\dtu,v_1+v_2)\Delta^2(v_1,v_2)\wp^{\prime 2}(\dtu)}
{\Delta^4(\dtu,v_1)}.
\ee
By means of the identity (\ref{wptwo}) $\wp(v_1+v_2) = \wp(2v_1)$ we find that
$\nu_1\nu_2=\nu_3\nu_4$ holds.

Thus we define the angle $\gamma(\dtu)$
\be
\ex{\ie\gamma(\dtu)} := \frac{\nu_2}{\nu_3} = \frac{\nu_4}{\nu_1}. \label{gamma}
\ee
If now $\chi_0(\dtu)=\gamma(\dtu)$ then the length $2\ell$ of the chord does not
depend on the arc $u$. If $\chi_0(\dtu)\not=\gamma(\dtu)$, then we may
introduce two of the curves with equal parameters $r_0$, $\epsilon$ and $\mu$.
If we rotate these two curves by the angle $\gamma-\chi_0$ against each other,
then again the chord connecting the points with arc parameter $u+\du$ on
one curve and arc parameter $u-\du$ on the other curve has for fixed $\du$
constant length $2\ell$. This property for infinitesimal $\ell$ allowed
initially the derivation of the differential equation (\ref{diffc}).

\subsection*{Increase of $\gamma$ with $\dtu$}

To investigate the variation of $\gamma$ let us factor $\ex{\ie\gamma}$
\bea
\ex{\ie\gamma(\dtu)} &=& \frac{\nu_4}{\nu_1} = \frac{n_0 n_1}{n^2_2 n_3},
\label{exigam} \\
n_0 &=& \frac{\Delta(\dtu,v_1)\Delta(\dtu,v_2)}{\wp^2(\dtu)\Delta(v_1,v_2)}, \\
n_1 &=& \Delta(\dtu+v_1,\dtu+v_2), \\
n_2 &=& \frac{\Delta(\dtu+v_1,\dtu)}{\wp(\dtu)}, \\
n_3 &=& \frac{\Delta(2\dtu+v_1,v_1)}{\Delta(\dtu+v_1,\dtu-v_1)}.
\eea
The first factor $n_0$ is positive and finite. Thus it does not contribute to
the variation of $\gamma$.

The factor $n_1$ varies from $n_1(0)=\wp(v_1)-\wp(v_2)>0$ to
$n_1(\omega_3)=\wp(v_1+\omega_3)-\wp(v_2+\omega_3)
=\wp(v_1+\omega')-\wp(v_2+\omega') =\wp(v'_1)-\wp(v'_2)<0$.
The imaginary part of $n_1$ is given by
\be
\Im n_1 = -\frac{\wp'(\dtu)}{2\ie} \left( \frac{\wp'(v_1)}
{(\wp(\dtu)-\wp(v_1))^2}
- \frac{\wp'(v_2)}{(\wp(\dtu)-\wp(v_2))^2} \right). \label{Imn1}
\ee
For large $\wp(\dtu)$ the second term in the large parenthesis is larger than
the first term, since $|\wp'(v_2)|>|\wp'(v_1)|$. Then $\Im n_1>0$.
The large parenthesis in (\ref{Imn1}) vanishes only for
\be
\frac{\wp(\dtu)-\wp(v_2)}{\wp(\dtu)-\wp(v_1)}=\pm\sqrt{\frac{\wp'(v_2)}
{\wp'(v_1)}}.
\ee
The left hand side increases from 1 at $\dtu=0$ to
$(e_3-\wp(v_2))/(e_3-\wp(v_1))$ at $\dtu=\omega_3$. Thus, if
\be
\frac{e_3-\wp(v_2)}{e_3-\wp(v_1)} = \frac{(1+\epsilon)^2}{(1-\epsilon)^2}
< \sqrt{\frac{\wp'(v_2)}{\wp'(v_1)}}
= \sqrt{\frac{(1+\epsilon)(1+\mu^2)}{(1-\epsilon)^3}},
\ee
which is the case for curves which obey
\be
\frac{(1+\epsilon)^3}{1-\epsilon} < 1+\mu^2,
\ee
then $\Im n_1>0$ in the whole interval. However, if $\epsilon>0$ and $\mu$ are
varied, then $\Im n_1(\dtu)>0$ will still hold at $\Re n_1(\dtu)=0$, since there
is no real solution $\dtu$ of $\wp(\dtu+v_1)=\wp(\dtu+v_2)$. The solution of
this equation reads $\dtu=\omega-(v_1+v_2)/2=\omega-v'_1$, where $\omega$ is a
period or half-period. All these solutions have an imaginary part. Therefore
variation of $\epsilon$ and $\mu$ will never yield a vanishing
$\Delta(\dtu+v_1,\dtu+v_2)$ for real $\dtu$. Thus $n_1$ contributes a change
$+\pi$ to $\gamma(\omega_3)-\gamma(0)$.

Next we consider the factor $n_2(\dtu)$. We find
\be
n_2(0)=-1, \quad n_2(\omega_3)=\frac{\wp(\omega_3+v_1)-\wp(\omega_3)}{e_3}
=\frac{\wp(v'_1)-e_3}{e_3} <0.
\ee
Here we have used $\wp(\omega_3+v_1)=\wp(\omega'+v_1)=\wp(\omega'-v_1)=\wp(v'_1)
<e_3$. The imaginary part of $n_2$ is given by
\be
\Im n_2(\dtu) = -\frac{\wp'(\dtu)\wp'(v_1)}{2\ie\wp(\dtu)
(\wp(\dtu)-\wp(v_1))^2}.
\ee
It is positive in the whole interval. Since $n_2$ has the same sign at both
endpoints, it does not contribute to $\gamma(\omega_3)-\gamma(0)$.

Finally we consider the factor $n_3(\dtu)$. At the endpoints of the interval
numerator and denominator vanish. The limits are
\be
n_3(0) = \frac{2\wp'(v_1)}{2\wp'(v_1)} =1, \quad
n_3(\omega_3) = \frac{2\wp'(2\omega_3+v_1)}{2\wp'(\omega_3+v_1)}
=\frac{\wp'(v_1)}{\wp'(\omega'+v_1)}=-\frac{\wp'(v_1)}{\wp'(v'_1)} <0.
\ee
The denominator of $n_3$ is purely imaginary
\be
\Delta(\dtu+v_1,\dtu-v_1) = -\frac{\wp'(\dtu)\wp'(v_1)}{\Delta^2(\dtu,v_1)}.
\ee
The numerator yields
\bea
\Delta(2\dtu+v_1,v_1) &=& R+\ie I, \\
R &=& -\wp(2\dtu)-2\wp(v_1)
+\frac{\wp^{\prime 2}(2\dtu)+\wp^{\prime 2}(v_1)}{4\Delta^2(2\dtu,v_1)}, \\
I &=& \ie \frac{\wp'(2\dtu)\wp'(v_1)}{2\Delta^2(2\dtu,v_1)}.
\eea
Thus the real part of $n_3$ is
\be
\Re n_3 = \frac{\wp'(2\dtu)}{2\wp'(\dtu)}
\frac{\Delta^2(\dtu,v_1)}{\Delta^2(2\dtu,v_1)}.
\ee
The second fraction is always positive. The first fraction is positive for
$0<\dtu<\omega_3/2$ and negative for $\omega_3/2<\dtu<\omega_3$. Thus the real
part turns from positive to negative at
$\dtu=\omega_3$. The imaginary part at $\dtu=\omega_3$ is given by
\be
\Im n_3 = \ie\frac{R(\omega_3/2)(\wp(\omega_3/2)-\wp(v_1))^2}
{\wp'(\omega_3/2)\wp'(v_1)}
\ee
Since $R(\omega_3/2)$ is negative ($\wp^{\prime 2}(v_1)$ is negative) and
$\wp'(\omega_3/2)$ is also negative, the sign of $\Im n_3$ is given by the sign
of $\ie/\wp'(v_1)$ which is negative.
Thus $n_3$ contributes $\pi$ to $\gamma(\omega_3)-\gamma(0)$ since $n_3$ appears
in the denominator of (\ref{exigam}).

In total we have
\be
\gamma(\omega_3)-\gamma(0) = 2\pi.
\ee
The condition for a constant length $\ell$ independent of $\tu$ is according to
(\ref{gamma})
\be
\nu_1=\nu_4 \ex{-\ie\chi_0}.
\ee
Thus the condition is
\be
\gamma(\dtu)-\chi_0(\dtu) \equiv 0 \,{\rm mod}\, 2\pi.
\ee
Thus any time $\gamma(\dtu)-\chi_0(\dtu)$ increases by $2\pi$ we have again a
chord of constant length $2\ell$. If one of the endpoints of the chord goes
around a closed curve of dihedral symmetry group $D_p$, then $\dtu$ increases
by $p\omega_3$. Then $\gamma(p\omega_3)-\chi_0(p\omega_3)=2(p-1)\pi$.
Thus we find $p-1$ solutions $\dtu$, for which $\ell$ is constant.
However, one solution is the trivial solution $\dtu=0$,
$\ell=0$. Thus we have $p-2$ non-trivial solutions in agreement with the
argument in \cite{WegnerI,WegnerII,Wegner}. We cannot exclude, that there are
more than $p-2$ solutions,
since we have not shown, that $\gamma(\dtu)-\chi_0(\dtu)$ is a monotonously
increasing function of $\dtu$.

\section{The Linear Case\label{linear}}

\subsection{The curve}

In the limit $r_0\rightarrow \infty$ one obtains the linear case, that is a
curve oscillating around a straight line. In this limit $\epsilon$ and $\mu$
approach 0 and we require for fixed $\xi$ and $\eta$
\be
r_0\rightarrow \infty, \quad \epsilon= \frac{\eta}{r_0}, \quad
\mu=\frac{\xi}{r_0}, \quad y=r-r_0, \quad x=r_0\psi.
\ee
$AD$ and $BC$ become infinitesimal with leading contributions
\be
AD=16\epsilon^2, \quad BC=16\mu^2.
\ee
Then the constants approach
\be
\lambda=\frac{\sqrt{\xi^2+\eta^2}}2, \quad
k=\frac{\eta}{2\lambda}, \quad
e_3=\frac{\xi^2-\eta^2}{24\lambda^2}.
\ee
The differential equations for the curve read
\bea
\frac 1{\sqrt{1+\left(\frac{\de y}{\de x}\right)^2}}
&=& \frac{2y^2+\xi^2-\eta^2}{4\lambda^2}, \\
\frac{\de u}{\de y} &=& \frac{2\lambda^2}{\sqrt{(\eta^2-y^2)(\xi^2+y^2)}}, \\
\frac{\de x}{\de u} &=& \frac{2y^2-\eta^2+\xi^2}{4\lambda^2}
\eea
One obtains $y$ as a function of the length of the arc $u=\lambda\tu$
\be
y=\eta \frac{-\wp(\tu)+e_3+\frac 14}{\wp(\tu)-e_3+\frac 14}. \label{y}
\ee
The solution for $x(\tu)$ is obtained from the solution for $\psi$.
It turns out, that $\tau$ becomes infinitesimal
\be
\tau = \frac{\epsilon}{2k}.
\ee
Therefore we can put
\be
v_i=v'_i=\frac{\omega'}2, \quad \wp(v_i)=\wp(v'_i) = e_3-\frac 14, \quad
\wp'(v_i)=\wp'(v'_i)=-\frac{\ie k}2,
\ee
unless we have to evaluate the difference of expressions, which differ only in
arguments $v$s and $v'$s, resp.
From(\ref{psiu}, \ref{c1c2}, \ref{c0}) we obtain
\be
x=\hat c_0 \tu +2\lambda (\zeta(\tu+v)+\zeta(\tu-v)) \label{x}
\ee
with
\be
\hat c_0 = \frac{\xi^2+2\eta^2}{3\lambda}.
\ee
By means of (\ref{zetaadd2}) we may rewrite
\be
\zeta(\tu+v)+\zeta(\tu-v) = 2\zeta(\tu)
+ \frac{\wp'(\tu)}{\wp(\tu)-e_3+\frac 14}.
\ee
During one period $x$ increases by
\be
x_{\rm per} = x(\tu+2\omega_3)-x(\tu) = 2\hat c_0\omega_3
+8\lambda \zeta(\omega_3). \label{xper}
\ee
Let us consider a few symmetries. First one sees immediately, that since we
start at the minimum $y(0)=-\eta$, $x$ is an odd and $y$ is an even function
of $\tu$,
\be
x(-\tu) = - x(\tu), \quad y(-\tu) = y(\tu).
\ee
Secondly if one increases $\tu$ by the half-period $\omega_3$, then $x$ is
shifted by $x_{\rm per}/2$ and $y$ changes sign,
\be
x(\tu+\omega_3) = x(\tu) + \frac{x_{\rm per}}2, \quad
y(\tu+\omega_3) = -y(\tu).
\ee
The result for $y$ can be seen by substituting
\be
\wp(\tu+\omega_3)=e_3 + \frac 1{16(\wp(\tu)-e_3)}. \label{wppomega3}
\ee
The result for $x$ is obtained by starting from (\ref{x}) and observing
\bea
\zeta(\tu+v+\omega_3) &=& \zeta(\tu-v+(\omega_3+\omega'))
=\zeta(\tu-v) -2\zeta(\omega_1), \\
\zeta(\tu-v+\omega_3) &=& \zeta(\tu+v) -2\zeta(\omega_2), \\
-2\zeta(\omega_1)-2\zeta(\omega_2) &=& 2\zeta(\omega_3).
\eea

\subsection{Length of the chord}

Again we consider the chord length $2\ell$ between the two points at $\tu+\dtu$
and $\tu-\dtu$. We are interested in the $\dtu$ for which the chord length
becomes independent of $\tu$. Therefore we determine the differences of the
coordinates. One obtains for the difference of the $x$-coordinates
\bea
\hat x &:=& x(\tu+\dtu)-x(\tu-\dtu) = 2\hat c_0 \dtu \\
&& + 2\lambda(\zeta(\tu+\dtu+v) + \zeta(\tu+\dtu-v) +\zeta(\dtu+v-\tu)
+\zeta(\dtu-v-\tu)). \nonumber
\eea
This can be rearranged to
\bea
\hat x &=& x_0(\dtu)
+2\lambda\left(\frac{\wp'(\dtu+v)}{\Delta(\dtu+v,\tu)}
+ \frac{\wp'(\dtu-v)}{\Delta(\dtu-v,\tu)}\right), \\
x_0(\dtu) &=& 2\hat c_0\dtu + 4\lambda (\zeta(\dtu+v) + \zeta(\dtu-v)).
\eea
We bring the difference of the $y$-coordinates to a common denominator
\be
\hat y := y(\tu+\dtu)-y(\tu-\dtu)
= -\frac{\eta}2 \frac{\Delta(\tu+\dtu,\tu-\dtu)}
{\Delta(\tu+\dtu,v)\Delta(\tu-\dtu,v)}.
\ee
We rewrite numerator and denominator
\bea
\Delta(\tu+\dtu,\tu-\dtu) &=& - \frac{\wp'(\tu)\wp'(\dtu)}{\Delta^2(\tu,\dtu)},
\\
\Delta(\tu+\dtu,v)\Delta(\tu-\dtu,v)
&=& \frac{\Delta(\tu,\dtu+v)\Delta(\tu,\dtu-v)\Delta^2(\dtu,v)}
{\Delta^2(\tu,\dtu)}.
\eea
and thus obtain
\be
\hat y = \frac{\eta}2 \frac{\wp'(\tu)\wp'(\dtu)}
{\Delta(\tu,\dtu+v)\Delta(\tu,\dtu-v)\Delta^2(\dtu,v)}.
\ee
Performing a partial fraction decomposition with respect $\wp(\tu)$ yields
\be
\hat y = \frac{\eta\wp'(\tu)}{2\wp'(v)}
\left(\frac 1{\Delta(\tu,\dtu+v)} -\frac 1{\Delta(\tu,\dtu-v)} \right).
\ee
In total we have
\bea
\hat x &=& x_0 +\frac{x_1}{\Delta(\tu,\dtu+v)} +\frac{x_2}{\Delta(\tu,\dtu-v)},
\\
x_1 &=& -2\lambda \wp'(\dtu+v), \quad x_2 = -2\lambda \wp'(\dtu-v), \\
\hat y &=& \frac{y_1\wp'(\tu)}{\Delta(\tu,\dtu+v)}
+ \frac{y_2\wp'(\tu)}{\Delta(\tu,\dtu-v)}, \\
y_1 &=& \frac{\eta}{2\wp'(v)}, \quad y_2 = - \frac{\eta}{2\wp'(v)}.
\eea
Thus the square of the length $2\ell$ of the chord can be written
\bea
4\ell^2 &=& \hat x^2 + \hat y^2
= x_0^2 + \frac{2x_0x_1}{\Delta(\tu,\dtu+v)}
+ \frac{2x_0x_2}{\Delta(\tu,\dtu-v)} \\
&+& \frac{x_1^2+y_1^2\wp^{\prime 2}(\tu)}{\Delta^2(\tu,\dtu+v)}
+ \frac{x_2^2+y_2^2\wp^{\prime 2}(\tu)}{\Delta^2(\tu,\dtu-v)}
+ \frac{2x_1x_2+2y_1y_2\wp^{\prime 2}(\tu)}
{\Delta(\tu,\dtu+v)\Delta(\tu,\dtu-v)}. \nonumber
\eea
With
\be
x_1^2+y_1^2\wp^{\prime 2}(\tu)
= 4\lambda^2 (\wp^{\prime 2}(\dtu+v)-\wp^{\prime 2}(\tu))
\ee
one obtains
\bea
\frac{x_1^2+y_1^2\wp^{\prime 2}(\tu)}{\Delta^2(\tu,\dtu+v)}
=-\frac{4\lambda^2(4\wp^2(\dtu+v)+4\wp(\dtu+v)\wp(\tu)+4\wp^2(\tu)-g_2)}
{\Delta(\tu,\dtu+v)} \nn
=-\frac{4\lambda^2(12\wp^2(\dtu+v)-g_2)}{\Delta(\tu,\dtu+v)}
-16\lambda^2(\wp(\tu)+2\wp(\dtu+v)).
\eea
A similar result holds for $(x_2^2+y_2^2\wp^{\prime 2}(\tu))/
\Delta^2(\tu,\dtu-v)$. Further we obtain
\be
x_1x_2+y_1y_2\wp^{\prime 2}(\tu) = 4\lambda^2 (\wp'(\dtu+v)\wp'(\dtu-v)
+\wp^{\prime 2}(\tu)),
\ee
which yields
\bea
\frac{2x_1x_2+2y_1y_2\wp^{\prime 2}(\tu)}{\Delta(\tu,\dtu+v)\Delta(\tu,\dtu-v)}
&=&32\lambda^2(\wp(\tu)+\wp(\dtu+v)+\wp(\dtu-v))
\\
&+& 8\lambda^2\frac{\wp'(\dtu+v)(\wp'(\dtu+v)+\wp'(\dtu-v))}
{\Delta(\tu,\dtu+v)\Delta(\dtu+v,\dtu-v)} \nn
&-& 8\lambda^2\frac{\wp'(\dtu-v)(\wp'(\dtu+v)+\wp'(\dtu-v))}
{\Delta(\tu,\dtu-v)\Delta(\dtu+v,\dtu-v)}. \nonumber
\eea
Collecting all terms $4\ell^2$ can be brought to the form
\bea
4\ell^2 &=& x_0^2 + \frac{\hat z_+}{\Delta(\tu,\dtu+v)}
+ \frac{\hat z_-}{\Delta(\tu,\dtu-v)}, \\
\hat z_+ &=& 2x_0x_1 -4\lambda^2 (12\wp^2(\dtu+v)-g_2) \nn
&+& 8\lambda^2 \frac{\wp'(\dtu+v)(\wp'(\dtu+v)+\wp'(\dtu-v))}
{\Delta(\dtu+v,\dtu-v)}, \\
\hat z_- &=& 2x_0x_2 -4\lambda^2 (12\wp^2(\dtu-v)-g_2) \nn
&-& 8\lambda^2\frac{\wp'(\dtu-v)(\wp'(\dtu+v)+\wp'(\dtu-v))}
{\Delta(\dtu+v,\dtu-v)}.
\eea
Since $v=\omega'/2$ one finds
\be
\hat z_+= 2x_1(x_0(\dtu) -d(\dtu)), \quad
\hat z_-= 2x_2(x_0(\dtu) -d(\dtu))
\ee
with
\bea
d(\dtu) &=&- \frac{\lambda k^2}4 \frac{\wp'(\dtu)}{\wp'(\dtu+v)\wp'(\dtu-v)}
\frac{\wp(\dtu)-e_3-\frac 14}{(\wp(\dtu)-e_3+\frac 14)^3} \nn
&=& -\lambda \frac{\wp'(\dtu)}{(\wp(\dtu)-e_3-\frac 14)(\wp(\dtu)-e_3+\frac 14)}.
\eea
The function $d(\dtu)$ has period $\omega_3$,
\be
d(\dtu+\omega_3) = d(\dtu).
\ee
One can see this by means of eq. (\ref{wppomega3}) and
\be
\wp'(\dtu+\omega_3) = -\frac{\wp'(\dtu)}{16(\wp(\dtu)-e_3)^2}
\ee
Referring to figure 2 the zeroes of $d$ are given by
\be
d(\rb{weierkr.eps}) =d(\rb{weierqu.eps}) =d(\rb{weierpl.eps})
=d(\rb{weiermi.eps}) = 0
\ee
and the poles by
\be
d(\rb{weierho.eps}) =d(\rb{weierbr.eps}) =\infty.
\ee
Thus along the real axis $d(\dtu)$ has poles at $\dtu=\omega_3(n+\frac 12)$
with integer $n$. With increasing argument it runs from $+\infty$ to $-\infty$.
Since $x_0(\dtu)$ is a slowly varying function without poles at finite $\dtu$,
there is in each interval $\omega_3(n-\frac 12) ... \omega_3(n+\frac 12)$ at
least one solution $\dtu$ for $x_0(\dtu)-d(\dtu)=0$, for which one obtains a
chord of constant length $x_0$.
Only the solution $\dtu=0$ is a trivial solution. Similarly as for the case of
a curve winding around the origin one can for given $\dtu$ find two copies of
the curve shifted by the distance $x_0-d$ along the x-axis, so that the chord
between these two curves stays constant, as one increases the arc parameter
on both curves by the same $u$.

\section{Shapes}

\subsection{Convexity}

If the curve bounds a convex region, then any chord between two points of the
curve will lie completely inside the region. Such a convexity is guaranteed,
if even at $r=r_0(1-\epsilon)$ the curvature is oriented to the center.
Therefore in the vicinity of this point one should have
$r=r_0(1-\epsilon)(1+\frac{h\psi^2}2)$ with $h<\frac 12$. One obtains
\be
f(r) = -\frac{\epsilon(1-\epsilon)((2-\epsilon)^2+\mu^2)}{\epsilon^2+\mu^2}
\frac{h\psi^2}2
= \frac{r^2}{\sqrt{r^2+r^{\prime 2}}} = -\frac{h^2}2(1-\epsilon)\psi^2
\ee
from which we conclude the condition for convexity
\be
h=\frac{2\epsilon((2-\epsilon)^2+\mu^2)}{\epsilon^2+\mu^2} \le \frac 12.
\ee
This condition is sufficient for a floating body, however, violation of
convexity is allowed for floating bodies, as long as the chord stays completely
inside the body.

\subsection{Radial and perpendicular tangents}

If $\epsilon$ ($\eta$) is not too large, then $\psi$ ($x$), resp. will be a
monotonously increasing functions of $\tu$. For sufficiently large $\epsilon$
however, the curves will bend over. Suppose this happens in the case of a curve
winding around the origin at $r=r_0\kappa$. Then the function $f(r)$,
eq. (\ref{fr}) obeys
\be
f(r_0\kappa) = -r_0\kappa,
\ee
from which we conclude
\be
(\kappa^2-1)^2 +(\kappa^2+1)\mu^2 -(\kappa^2-1)\epsilon^2 -\epsilon^2\mu^2 =0
\ee
with the solution
\be
\kappa^2 = 1+\frac{\epsilon^2-\mu^2}2
\pm\sqrt{2\epsilon^2-2\mu^2+\left(\frac{\epsilon^2+\mu^2}2\right)^2}.
\ee
Thus the condition
\be
2\epsilon^2-2\mu^2+\left(\frac{\epsilon^2+\mu^2}2\right)^2 \ge 0
\ee
has to be fulfilled in order to obtain a curve with radial tangent.
We note, that in this case there are two solutions $\kappa$ with
\be
\kappa_1\kappa_2 = \sqrt{(1-\epsilon^2)(1+\mu^2)}.
\ee
In the case of the linear curve one obtains perpendicular tangents at
\be
y_{\perp} = \pm \sqrt{\frac{\eta^2-\xi^2}2}
\ee
provided the condition $\eta^2\ge\xi^2$ is met.

Bracho, Montejano and Oliveros \cite{Oliveros,Bracho} have determined tire
track curves with the special restriction, that the arc between the two end
points of the chord is one fifth of the perimeter by introducing a carousel
consisting of an equilateral pentagon with the property, that the midpoints
of the sides move parallel to them. The midpoints describe the tracks $\gamma$
of the rear wheels. The corners of the pentagon describe the tracks
$\Gamma$ of the front wheels. They require that all five midpoints move on
the same curve and consequently the corners describe the same curves.
They have found closed
curves for $p=7$ (fig. 3 of \cite{Bracho} and fig. 9 of \cite{Oliveros})
and for $p=12$ (fig. 1 of \cite{Bracho}). I do not know, whether these curves
belong to the solutions given here, but I will assume this in the following
discussion. Consequently the curve with $p=7$ is
simultaneously a tire track curve $\Gamma$ for two other chord lengths.
Since $p$ is odd, one of the chord lenths belongs to perimeter ratio 1/2.
It is a special Auerbach curve\cite{Auerbach}, although it is not convex.
Similarly I expect that the curve for $p=12$ is a tire track curve for five
further chord lengths. We note, that the requirement of finding a closed
curve which winds once around the center and the requirement of spanning
one fifth of the
perimeter between the endpoints of the chord yields only discrete solutions
for $\epsilon$ and $mu$, so that one cannot require simultaneously
convexity. Since in our approach the fraction of the perimeter covered
by the arc between the endpoints of the chord can be varied continuously,
we can obtain convex solutions. A solution, which is not convex, but
'suffiently convex' to serve for a floating body for $p=7$ is shown in fig.
4 of \cite{WegnerII} and fig. 1 of \cite{Wegner}.

Fig. 2 of \cite{Bracho} shows a set of five curves. They have identical
shape, but are shifted against each other by equal distances. They can
be numbered 1 to 5 and have the property, that any pair of curves $i$ and $i+1$
($i+5\equiv i$) can be connected by a chord of constant length, as the ends of
the chord move by equal arc lengths. This is in agreement with the behavior
we described in the linear case. Here however, the authors required, that after
traversing through five chords, one does not only return to the initial curve
but even to the same point. (The corners of the equilateral pentagons
can be well seen at the left ends and at the right ends of the depicted parts
of the curves.) From the discussion in this paper I conclude,
that there is an infinity of chord lengths with this property, if one waives
the  requirement, that after traversing through five chords one returns to the
same point. Taking only one of these curves, it is also a tire track curve
for an infinity of chord lengths.

\subsection{Eights}

An extreme case is the situation, where a curve with tangential or
perpendicular slopes returns to the initial point
after an increase of $\tu$ by $2\omega_3$.
Then such a curve has the shape of an eight and $\psi_{\rm per}=0$
(eq. \ref{psiper}) and $x_{\rm per}=0$ (eq. \ref{xper}), resp., have to be
fulfilled. An example of an eight can be found in fig. 8 of \cite{Oliveros}.
\vspace{5mm}

\subsection{Other shapes}

In this paper I have assumed during the discussion of the curves, that $\mu$ is
real and tacitely $-1<\epsilon<1$. Obviously there are also other values of
$\epsilon$ and $\mu$ which yield real polar coordinates $r$ and $\psi$. The
properties of the corresponding curves will be investigated elsewhere.

\paragraph{Acknowledgment} I am indebted to Serge Tabachnikov for useful
correspondence.

\begin{appendix}
\section{Some Formulae}

\subsection{Some formulae for the Weierstrass function and its
integrals\label{formulae}}

The Weierstrass function $\wp$ is defined by
\be
\wp^{\prime 2}(z) = 4\wp^3(z)-g_2\wp(z)-g_3
= 4(\wp(z)-e_1)(\wp(z)-e_2)(\wp(z)-e_3) \label{diff}
\ee
with the requirement that one of the singularities is at $z=0$ and
\be
e_1+e_2+e_3=0.
\ee
Commonly the two integrals are defined
\bea
\zeta(a) &=& \frac 1a - \int_0^a (\wp(z)-\frac 1{z^2}) \de z, \label{zetaint} \\
\sigma(a) &=& a \exp\left(\int_0^a (\zeta(z)-\frac 1z) \de z\right),
\label{sigmaint}
\eea
The function $\wp$ is an even function of its argument, $\zeta$ and $\sigma$ are
odd functions. The Laurent and Taylor expansions start with
\bea
\wp(a) &=& \frac 1{a^2} + \frac{g_2}{20} a^2 + \frac{g_3}{28} a^4 + ... \\
\zeta(a) &=& \frac 1a - \frac{g_2}{60} a^3  - \frac{g_3}{140} a^5 - ... \\
\sigma(a) &=& a - \frac{g_2}{240} a^5 - \frac{g_3}{840} a^7 - ...
\eea
There exist addition theorems
\bea
\wp(a+b) &=& -\wp(a) -\wp(b) + \frac{(\wp'(a)-\wp'(b))^2}{4(\wp(a)-\wp(b))^2},
\label{wpadd}, \\
\zeta(a+b) &=& \zeta(a)+\zeta(b) +\frac 12 \frac{\wp'(a)-\wp'(b)}
{\wp(a)-\wp(b)}, \label{zetaadd} \\
\sigma(a+b)\sigma(a-b) &=& -\sigma^2(a)\sigma^2(b)(\wp(a)-\wp(b)),
\label{sigmaadd}
\eea
If $\omega$ is a half-period, that is $\omega$ itself is not a period of $\wp$,
but $2\omega$ is, then the following relations hold for integer $n$
\bea
\wp(a+2n\omega) &=& \wp(a), \label{wpper} \\
\zeta(a+2n\omega) &=& \zeta(a) + 2n\zeta(\omega), \label{zetaper} \\
\sigma(a+2n\omega) &=& (-)^n\sigma(a) \ex{2n(a+n\omega)\zeta(\omega)}.
\label{sigmaper}
\eea
From (\ref{zetaadd}) one obtains
\be
\zeta(a+b)+\zeta(a-b) = 2\zeta(a)+\frac{\wp'(a)}{\wp(a)-\wp(b)}.
\label{zetaadd2}
\ee

\subsection{cn in terms of the Weierstrass $\wp$}

We perform the transformations of Jacobi's elliptic functions $\cn$, $\dn$ and
$\sn$
\bea
\cn(\tu,k) &=& \dn(k\tu,\frac 1k), \\
\dn((1+k_1)u',\frac{2\sqrt{k_1}}{1+k_1})
&=& \frac{1-k_1\sn^2(u',k_1)}{1+k_1\sn^2(u',k_1)}
\eea
with
\bea
\frac{2\sqrt{k_1}}{1+k_1}=\frac 1k, && \sqrt{k_1} = k+\ie\sqrt{1-k^2}, \\
k_1^{\pm 1} = 2k^2-1 \pm 2\ie k\sqrt{1-k^2}, &&
u'=\frac{ku}{1+k_1} = \frac u{2\sqrt{k_1}}.
\eea
Note, that $k_1$ and $u'$ are complex. We express $\sn$ by the Weierstrass $\wp$
function
\bea
\sn(u',k_1) = \sn(\frac{\tu}{2\sqrt{k_1}},k_1) =
\frac{\sqrt{e_1-e_3}}{\sqrt{\wp(\tu)-e_3}}, \\
k_1^2=\frac{e_2-e_3}{e_1-e_3}, \quad \sqrt{e_1-e_3}=\frac 1{2\sqrt{k_1}},
\quad e_1+e_2+e_3=0.
\eea
Then we obtain
\bea
e_1 &=& \frac 1{12k_1}(2-k_1^2) = \frac 1{12}(2k^2-1-6\ie k\sqrt{1-k^2}),
\label{e1}\\
e_2 &=& \frac 1{12k_1}(-1+2k_1^2) = \frac 1{12}(2k^2-1+6\ie k\sqrt{1-k^2}),
\label{e2}\\
e_3 &=& \frac 1{12k_1}(-1-k_1^2) = \frac 16(1-2k^2), \\
g_2 &=& -4(e_1e_2+e_1e_3+e_2e_3) = -\frac 14+\frac 13(1-2k^2)^2
= -\frac 14 +12 e_3^2, \label{g2} \\
g_3 &=& 4e_1e_2e_3 = \frac 1{24}(1-2k^2) -\frac 1{27}(1-2k^2)^3
= \frac 14 e_3 - 8 e_3^3. \label{g3}
\eea
Thus we obtain finally for $\cn$
\be
\cn(\tu,k) = \frac{\wp(\tu,g_2,g_3)-e_3-\frac 14}
{\wp(\tu,g_2,g_3)-e_3+\frac 14}
\ee
and for the derivative of $\wp$ from (\ref{diff})
\be
\wp^{\prime2}(\tu) = 4(\wp(\tu)-e_3)
(\wp(\tu)^2 +e_3 \wp(\tu)+\frac 1{16}-2e_3^2).
\label{wps}
\ee

\subsection{Three-Arguments Addition Theorem}

Using the addition theorem (\ref{wpadd}) one finds an addition theorem including
three arguments
\bea
&& (\wp(a+b)-\wp(c))(\wp(a-b)-\wp(c))(\wp(a)-\wp(b))^2 \nn
&=& (\wp(a+c)-\wp(b))(\wp(a-c)-\wp(b))(\wp(a)-\wp(c))^2 \nn
&=& (\wp(b+c)-\wp(a))(\wp(b-c)-\wp(a))(\wp(b)-\wp(c))^2.
\label{add2}
\eea
In the following its derivation is sketched. Using (\ref{wpadd}) one obtains
\bea
&& (\wp(a+b)-\wp(c))(\wp(a-b)-\wp(c))(\wp(a)-\wp(b))^2 \nn &=&
\left(-S+\frac{(\wp'(a)-\wp'(b))^2}{4(\wp(a)-\wp(b))^2}\right)
\left(-S+\frac{(\wp'(a)+\wp'(b))^2}{4(\wp(a)-\wp(b))^2}\right)(\wp(a)-\wp(b))^2
\nn
&=& S^2(\wp(a)-\wp(b))^2 -\frac S2 (\wp^{\prime 2}(a)+\wp^{\prime 2}(b))
+\frac{(\wp^{\prime 2}(a)-\wp^{\prime 2}(b))^2}{16(\wp(a)-\wp(b))^2}
\eea
with
\be
S=\wp(a)+\wp(b)+\wp(c).
\ee
One inserts the expressions for $\wp^{\prime 2}$ and obtains for the last term
\be
\frac{(\wp^{\prime 2}(a)-\wp^{\prime 2}(b))^2}{16(\wp(a)-\wp(b))^2}
=(\wp^2(a)+\wp(a)\wp(b)+\wp^2(b)-\frac 14g_2)^2.
\ee
Now adding all contributions one obtains
\bea
&& (\wp(a+b)-\wp(c))(\wp(a-b)-\wp(c))(\wp(a)-\wp(b))^2 \nn
&=& \wp^2(a)\wp^2(b)+\wp^2(a)\wp^2(c)+\wp^2(b)\wp^2(c) \nn
&-& 2(\wp(a)+\wp(b)+\wp(c))\wp(a)\wp(b)\wp(c) \nn
&+& \frac{g_2}2 (\wp(a)\wp(b)+\wp(a)\wp(c)+\wp(b)\wp(c)) \nn
&+& g_3(\wp(a)+\wp(b)+\wp(c))+\frac{g_2^2}{16}.
\eea
This expression is invariant under any permutations of $a$, $b$, and $c$. Thus
(\ref{add2}) follows.

\subsection{Another Three-Arguments Addition Theorem}

Starting from the addition theorem (\ref{wpadd}) we obtain
\be
\Delta(a\pm b,a\pm c) = \Delta(c,b)
+\frac 14\left(\frac{\wp'(a)\mp\wp'(b)}{\Delta(a,b)}\right)^2
-\frac 14\left(\frac{\wp'(a)\mp\wp'(c)}{\Delta(a,c)}\right)^2.
\ee
Thus we obtain
\bea
&& \Delta(a+b,a+c)\Delta(a-b,a-c) \nn
&=& \left(\Delta(c,b)
+\frac 14 \frac{\wp^{\prime 2}(a)+\wp^{\prime 2}(b)}{\Delta^2(a,b)}
-\frac 14 \frac{\wp^{\prime 2}(a)+\wp^{\prime 2}(c)}{\Delta^2(a,c)} \right)^2
\nn
&-& \frac 14\left(\frac{\wp'(a)\wp'(b)}{\Delta^2(a,b)}
-\frac{\wp'(a)\wp'(c)}{\Delta^2(a,c)}\right)^2.
\eea
Some rearrangement yields
\bea
&& \Delta(a+b,a+c)\Delta(a-b,a-c) = A^2 + \wp^{\prime 2}(a) B, \label{prod} \\
&& A =\Delta(c,b)
- \frac 14 \frac{\wp^{\prime 2}(a)-\wp^{\prime 2}(b)}{\Delta^2(a,b)}
+ \frac 14 \frac{\wp^{\prime 2}(a)-\wp^{\prime 2}(c)}{\Delta^2(a,c)}, \\
&& B = \Delta(c,b) \left(\frac 1{\Delta^2(a,b)} - \frac 1{\Delta^2(a,c)}\right)
-\frac 14 \frac{(\wp'(b)-\wp'(c))^2}{\Delta^2(a,b)\Delta^2(a,c)}.
\eea
Using
\be
\frac{\wp^{\prime 2}(a)-\wp^{\prime 2}(b)}{\Delta(a,b)}
=4\wp^2(a)+4\wp(a)\wp(b)+4\wp^2(b)-g_2
\ee
we obtain
\be
A= \frac{\Delta(c,b)}{\Delta(a,b)\Delta(a,c)}
\left(3\wp^2(a)-\frac{g_2}4\right).
\ee
For $B$ we obtain
\be
B= -\frac{\Delta^2(c,b)}{\Delta^2(a,b)\Delta^2(a,c)} (2\wp(a)+\wp(b+c)).
\ee
Substitution of $A$ and $B$ into (\ref{prod}) yields
\bea
&& \Delta(a+b,a+c)\Delta(a-b,a-c) = \nn
&& \frac{\Delta^2(c,b)}{\Delta^2(a,b)\Delta^2(a,c)}
\left(\wp^4(a)+\frac{g_2}2\wp^2(a)+2g_3\wp(a)+\frac{g_2^2}{16}
- \wp^{\prime 2}(a)\wp(b+c)\right). \nn &&
\eea
Since
\be
\wp^4(a) + \frac{g_2}2 \wp^2(a) +2g_3\wp(a) +\frac{g_2^2}{16}
= \wp(2a)\wp^{\prime 2}(a),
\ee
we obtain finally
\be
\Delta(a+b,a+c)\Delta(a-b,a-c)
= \frac{\Delta^2(c,b)\Delta(2a,b+c)\wp^{\prime 2}(a)}
{\Delta^2(a,b)\Delta^2(a,c)}. \label{add4}
\ee
In the limit $c\rightarrow 0$ we obtain
\be
\Delta(a+b,a)\Delta(a-b,a)
= \frac{\Delta(2a,b)\wp^{\prime 2}(a)}{\Delta^2(a,b)}.
\label{add5}
\ee
Further we obtain
\be
\Delta(a+b,a-b)
=\frac 14 \frac{(\wp'(a)-\wp'(b))^2}{\Delta^2(a,b)}
-\frac 14 \frac{(\wp'(a)+\wp'(b))^2}{\Delta^2(a,b)}
=-\frac{\wp'(a)\wp'(b)}{\Delta^2(a,b)}. \label{add6}
\ee
\end{appendix}

\end{document}